# Galaxies appear simpler than expected

M. J. Disney[1], J. D. Romano[1,2], D. A. Garcia-Appadoo[3,1], A. A. West[4,5], J. J. Dalcanton[5]
& L. Cortese[1]

[1]*School of Physics & Astronomy, Cardiff University, 5 The Parade, Cardiff CF24 3AA,
UK*

[2]*Department of Physics and Astronomy, The University of Texas at Brownsville, 80 Fort
Brown, Brownsville, TX 78520, USA*

[3]*European Southern Observatory, Alonso de Cordova 3107,Casilla 19001, Vitacura,
Santiago 19, Chile*

[4]*Astronomy Department, University of California, 601 Campbell Hall, Berkeley, CA
94720-3411, USA*

[5]*Department of Astronomy, Physics-Astronomy Building C309, University of
Washington, Box 351580, Seattle, WA 98195, USA*

**Galaxies are complex systems the evolution of which apparently results from
the interplay of dynamics, star formation, chemical enrichment, and feedback
from supernova explosions and supermassive black holes[1]. The hierarchical
theory of galaxy formation holds that galaxies are assembled from smaller pieces,
through numerous mergers of cold dark matter[2,3,4]. The properties of an
individual galaxy should be controlled by six independent parameters including
mass, angular-momentum, baryon-fraction, age and size, as well as by the
accidents of its recent haphazard merger history. Here we report that a sample of
galaxies that were first detected through their neutral hydrogen radio-frequency
emission, and are thus free of optical selection effects[5], shows five independent
correlations among six independent observables, despite having a wide range of**

**properties. This implies that the structure of these galaxies must be controlled by a single parameter, although we cannot identify this parameter from our dataset. Such a degree of organisation appears to be at odds with hierarchical galaxy formation, a central tenet of the cold dark matter paradigm in cosmology[6].**

About 300 sources, from part of the much larger blind 21-cm survey for neutral hydrogen made with the Parkes radio telescope[7,8,9], overlap a region surveyed by the Sloan Digital Sky Survey (SDSS) in the optical spectral region[10]. Two hundred were unambiguously identified as individual galaxies, and are representative of the whole range of galaxies between giant spirals and extreme dwarfs, missing only the ~10% of largely neutral-gas-free galaxies found mainly in big clusters. For each object, we measured the HI (neutral hydrogen) mass and line width $\Delta V$, redshift, inclination, two radii ($R_{50}$ and $R_{90}$) respectively containing 50 and 90% of the emitted light, the luminosity $L_g$, and 4 colours. Three of the colours are degenerate[11,12], leaving dynamical mass ($M_d \equiv (\Delta V)^2 R_{90}/G$, ref. 13), HI mass ($M_{HI}$), luminosity, radius, concentration ($R_{50}/R_{90}$), and colour (SDSS ($g$-$r$)). For comparison, we can imagine galaxies being controlled by seven physical quantities, namely total mass, baryon fraction, age, specific angular momentum, specific heat energy (random motion), radius and concentration, only six of which can be independent, owing to the Virial theorem. We thus have as many independent observables as we do controlling physical parameters, making an examination of the correlation structure potentially very interesting. Each discovered correlation, if independent of the rest, will set a further constraint on galaxy physics.

We find five correlations, four of which were known already: that between dynamical mass and luminosity ($M_d \propto L_g$)[14], that between luminosity and colour (luminous galaxies are redder)[15], that between $R_{50}$ and $R_{90}$ (HI galaxies have exponential profiles)[16], and that between $M_{HI}$ and $R_{50}$ (all the galaxies have the same HI surface-density $M_{HI}/R_{50}^2$)[17,18,11]. The new correlation, namely that surface brightness $\Sigma \equiv$ $L_g/R_{50}^2$ is proportional to $R_{50}$ (ref. 11), required a blind HI survey to clearly separate it from the pronounced selection effects in the optical. All the data and the correlations appear in ref. 11.

Here we examine the correlation structure by means of a Principal Component Analysis (PCA)[19,20] based on the correlation matrix of the measured data. Being normalized (unlike the covariance matrix), a correlation-based analysis is immune to the influence of scaling. One can think of PCA as a search in the six-dimensional space of observables for a smaller number of coordinates that describe nearly all the variance. For instance, PCA has been used to show that elliptical galaxies lie on a 'fundamental plane'[21,22], that is, in a two-dimensional space. Because such a correlation analysis relies on linear relations between variables, we use logarithmic quantities (the colour, being a magnitude, is also logarithmic)[23].

Colour turns out to be more complex than the other observables, so we omit it at first then reintroduce it later. Figure 1 demonstrates the strong correlations that exist between the five other variables. This is emphasised in Fig. 2 where all are seen to be strongly correlated with the first principal component, PC1, and scattered with respect to the other principal components. A high degree of organisation is already evident. The eigenvalue of PC1 is 4.1, in comparison with a maximum possible of 5.0 (1 for each variable; they should sum to 5), while the next principal component has an eigenvalue of only 0.53. We note that all five correlations are independent because each introduces at least one *a priori* independent observable not present in any preceding correlation. We note further that each implies some new *intrinsic* (not merely scaling) property that calls for a physical explanation. For instance, the correlation between surface brightness and $R_{50}$ implies $L_g/R_{50}^2 \propto R_{50}$ or that the luminosity-density (defined as $L_g/R_{50}^3$) is independent of either luminosity or dynamical mass.

Figure 3 shows the correlations when the colour is included. Colour is evidently correlated with all the remaining variables, but more weakly so. (The least significant such correlation (0.39) is nevertheless significant at the 0.01% level). Figure 4 clarifies the situation by exhibiting the principal component structure. Like all the other variables, colour is definitely correlated with PC1, which now has an eigenvalue of 4.4 out of a maximum possible of 6.0 (that is, explains 73% of the variance). The next highest eigenvalue is 0.75 for PC2 (Fig. 4, column two). Statisticians normally ignore, as insignificant, principal components with eigenvalues of less than 1.0, except in the special case in which one observable is scattered mostly-independently of all the rest, when values as low as 0.7 may be considered significant[19]. This turns out to be the case here (0.75). There is a component of the colour that is correlated with nothing but itself, and this 'rogue' component constitutes a weak, but nevertheless significant, principal component, PC2, as is clear from column two. The strong correlation in the colour plot of column two is notable. Beyond the fact that PC2 is strongly aligned with colour (direction cosine 0.83) what does this mean? Simulations (Appendix A) show that if colour were perfectly uncorrelated with everything else (that is, was randomly scattered), then the colour plot of column two would be a sharp, straight line at a slope of 45 degrees. This is because even if an observable (for example colour) is correlated with nothing else it is nevertheless correlated perfectly with itself, and the PCA seeks out a principal component aligned perfectly with that observable (in this case colour).

We can summarise the PCA as follows: HI-selected galaxies appear to have colours made up of two components, which we label the 'systematic' component and the 'rogue' component. The rogue colour is scattered more or less randomly, and is correlated with none of the other observables. (It would be natural, although not compelling, to identify the systematic colour with the old population of stars, and the rogue component with recent and transitory bursts of star formation—very luminous young stars being far bluer and shorter lived than older ones.) The remaining six variables, including the systematic colour, are all correlated with one another, and with a single principal component. In other words they form a one-parameter set lying on a single fundamental line. Such simplicity was unpredicted and is noteworthy when galaxies might well have been controlled by any six of the seven potentially independent physical parameters listed earlier. It is even more significant considering the enormous variety amongst the galaxies in the sample[11].

If, as we have argued, galaxies come from at most a six-parameter set, then for gaseous galaxies to appear as a one-parameter set, as observed here, the theory of galaxy formation and evolution must supply five independent constraint equations to constrain the observations. This is such a stringent set of requirements that it is hard to imagine any theory, apart from the correct one, fulfilling them all. For instance, consider heirarchical galaxy formation in the dark matter model, which has been widely discussed in the literature[3,4]. Even after extensive simplification, it still contains four parameters per galaxy: mass, spin, halo-concentration index and epoch of formation. Consider spin alone, which is thought to be the result of early tidal torquing. Simulations produce spins, independent of mass, with a log-normal distribution. Higher-spin discs naturally cannot contract as far; thus, to a much greater extent than for low-spin discs, their dynamics is controlled by their dark halos, so it is unexpected to see the nearly constant dynamical-mass/luminosity ratio that we and others[14] actually observe. Heirarchical galaxy formation simply does not fit the constraints set by the correlation structure in the Equatorial Survey.

More generally, a process of hierarchical merging, in which the present properties of any galaxy are determined by the necessarily haphazard details of its last major mergers, hardly seems consistent with the very high degree of organisation revealed in this analysis. Hierarchical galaxy formation does not explain the commonplace gaseous galaxies we observe. So much organization, and a single controlling parameter—which cannot be identified for now—argue for some simpler model of formation. It would be illuminating to identify the single controlling galaxy parameter, but this cannot be attempted from the present data. To confuse correlation with cause is the critical mistake that can be made in this kind of analysis.

It is natural to ask why this fundamental line was not discovered before. To some extent it was because even the pioneers[24,25] and others[26,27,28] working with small numbers of optically selected spirals could reduce six observables to two; one relating to size, one to morphology. The strong optical selection effects, which hamper optical astronomers in detecting and measuring galaxies whose surface brightnesses are barely brighter than the sky[5,29], disguised galaxies' simplicity.

**Acknowledgements** We would like to thank the HIPASS team, and especially R. Ekers, A. Wright and L. Staveley-Smith of the Australian National Telescope at CSIRO Radiophysics in Sydney for their foresight and enterprise in getting the Multibeam project started. MJD would like to thank Mathias Disney of the Geography Department at University College London for first pointing out the one-dimensional nature of this data.

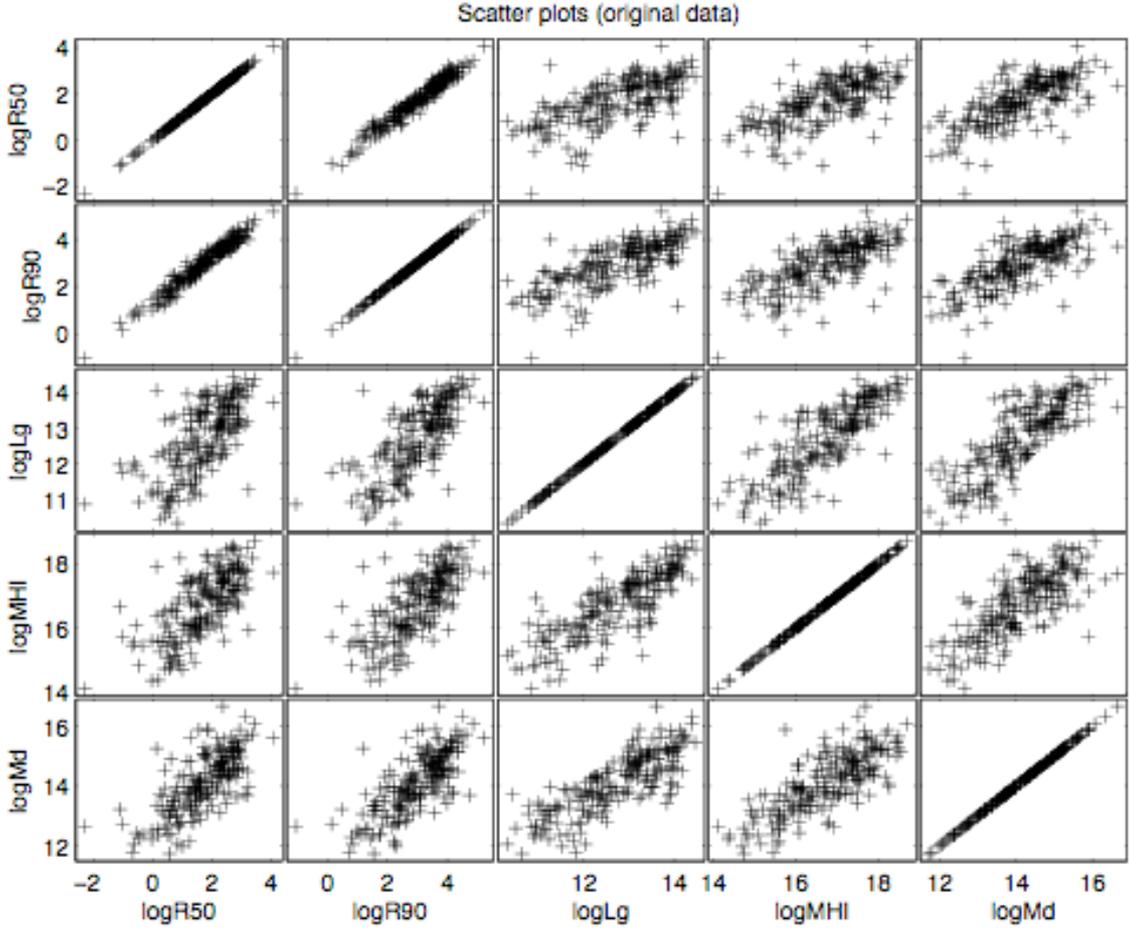

Figure 1: Scatter plots showing correlations between five measured variables, not including colour. The variables are two optical radii, $R_{50}$ and $R_{90}$ (in kiloparsecs), respectively containing 50 and 90% of the emitted light; luminosity, $L_g$; neutral hydrogen mass, $M_{HI}$; and dynamical mass, $M_d$ (inferred from the 21-cm linewidth, the radius and inclination in the usual way; see main text), all in solar units. In this normalized, logarithmic representation, the slopes are irrelevant, only the scatters are significant. We find that (see ref. 11): $R_{90} \propto R_{50}$, $L_g \propto R_{50}^3$, $M_{HI} \propto R_{50}^2$ and $M_d \propto L_g$. Alternatively, the surface brightness $\Sigma \equiv L_g/R_{50}^2 \propto L_g^{1/3} \propto R_{50}$, whereas the luminosity density (defined as $L_g/R_{50}^3$) and mass density (defined as $M_d/R_{50}^3$) are independent of size. Colour is included in Fig. 3. To see the same figures in more comprehensible, un-normalized units see Appendix B.

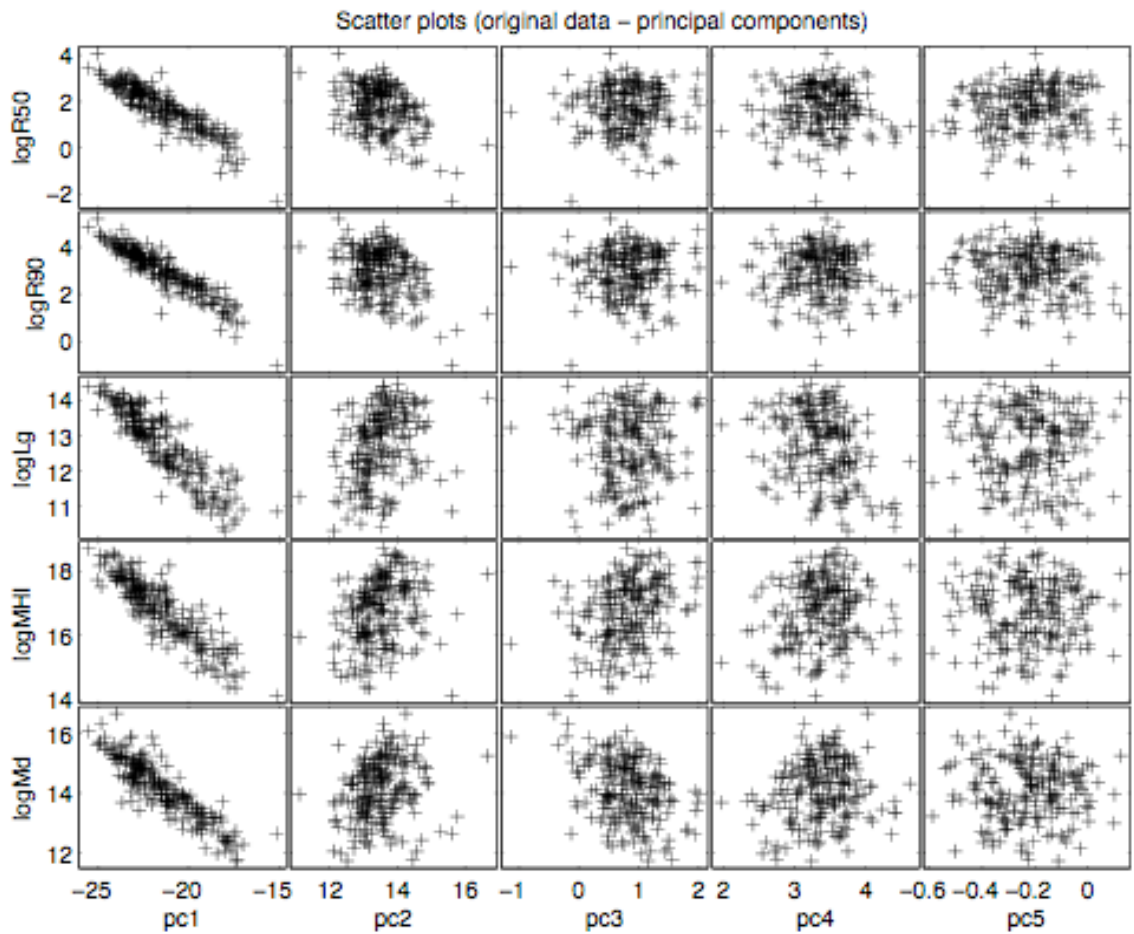

Figure 2: Scatter plots showing correlations between the five measured variables and the principal components. The five physical quantities are tightly correlated with a single principal component, PC1, which accounts for their correlations with one another. The principal components are labelled from the strongest, PC1, to the weakest, PC5, with eigenvalues 4.1, 0.53, 0.23, 0.17, and 0.02, respectively. Eigenvalues less than 1 are normally thought to represent scatter only—but see main text. The direction cosines for PC1 relative to the physical variables in order plotted (top to bottom) are -0.45, -0.46, -0.44, -0.44, -0.44 (that is, approximately -$1/\sqrt{5}$). This is expected for such a strong principal component and makes it impossible to single out any one observable as the 'driving force'. In any case, correlations alone should not be used to infer causation.

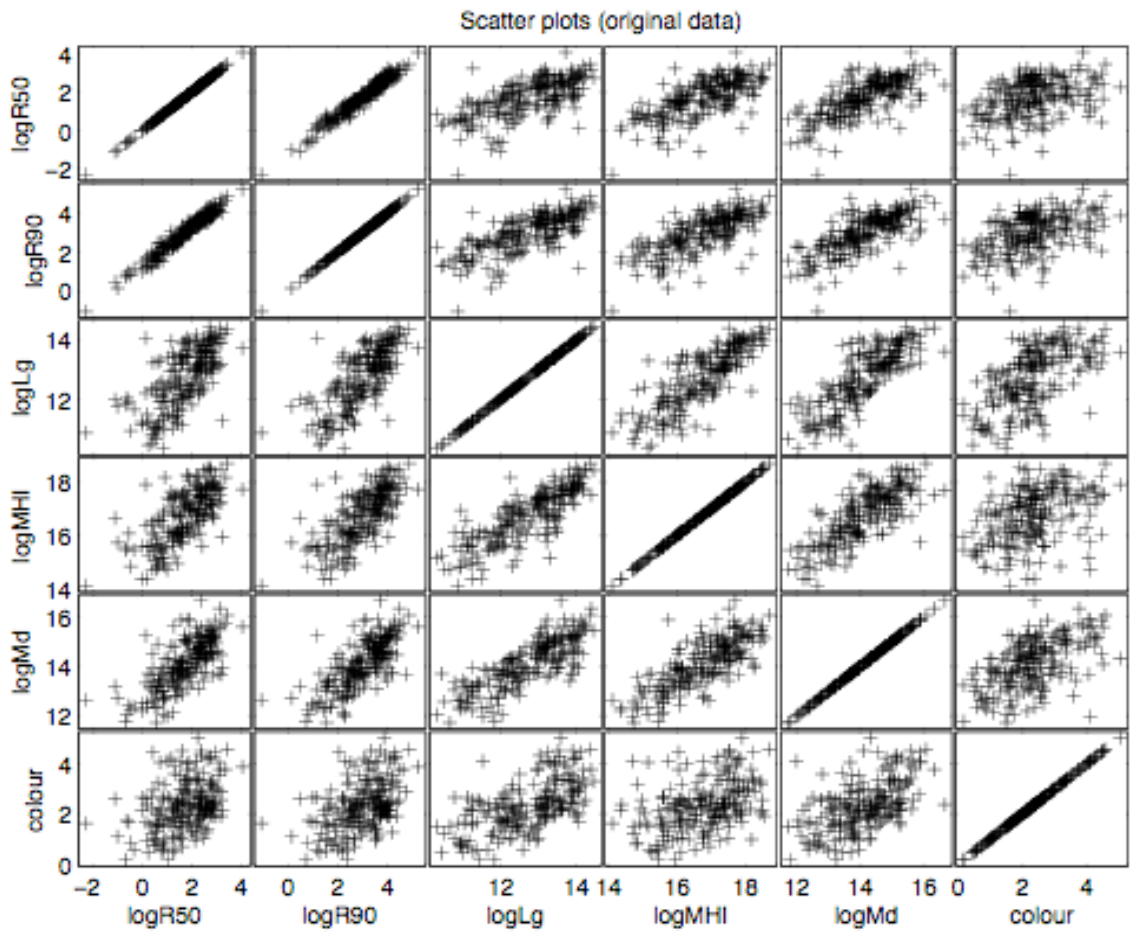

Figure 3: Scatter plots showing correlations between the previous five measured variables (Fig. 1), but now including colour. The colour is (g-r), that is, 'green minus red' and is also logarithmic. Colour (measured within $R_{90}$) is clearly correlated with the five other observables in the sense that more luminous galaxies are redder, which is usually taken to mean that their output is dominated by older stars. However, the colour correlations are much weaker, though still highly significant (at the 0.01% level).

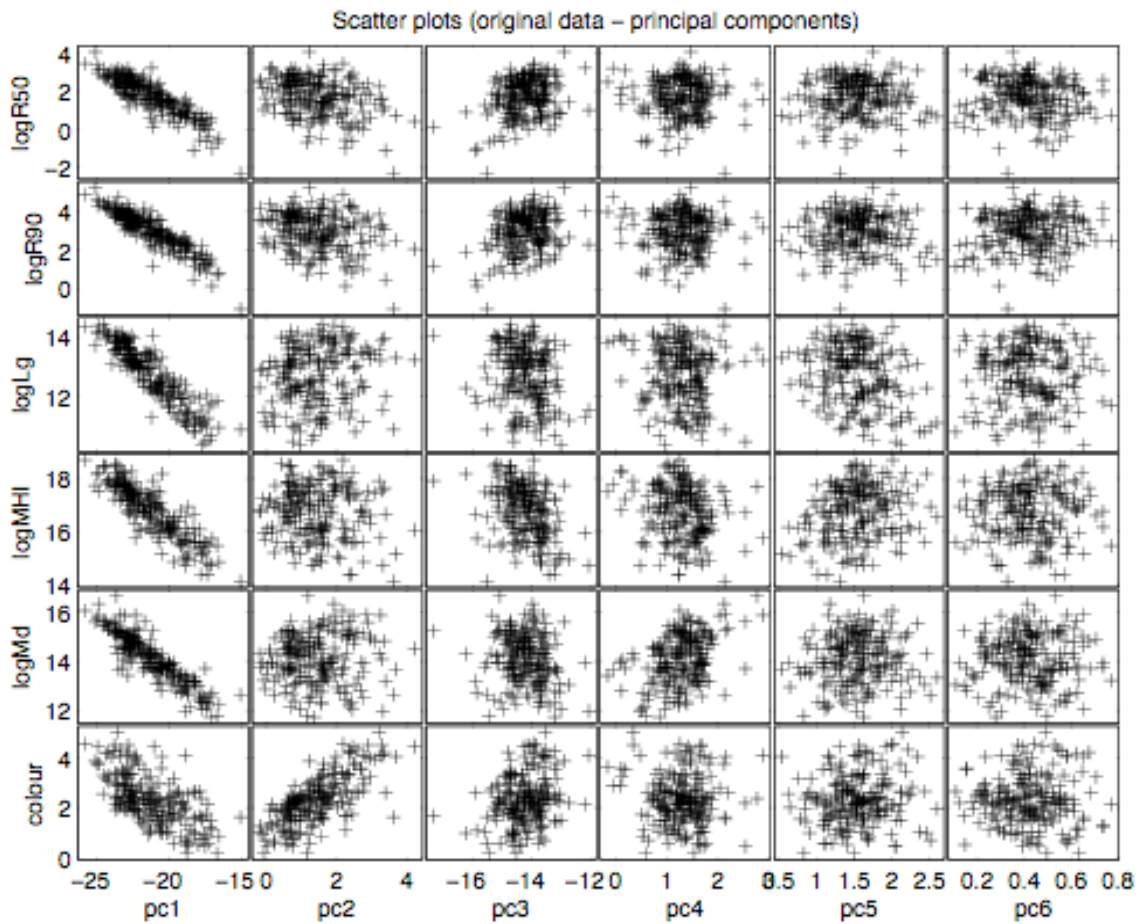

Figure 4: Scatter plots showing correlations between all the six measured variables (including colour) and the principal components. The principal components are labelled from the strongest, PC1, to the weakest, PC6, with eigenvalues 4.4, 0.75, 0.48, 0.22, 0.14, and 0.02, respectively. The bottom plot of column one shows that colour is well correlated with the other five observables and with PC1. However, the bottom plot in column two shows that colour is even more strongly-correlated with a new principal component, PC2, which correlates with nothing else. If colour were entirely independent of all the other variables, it would have its own principal component with an eigenvalue of 1.0, instead of 0.75 as observed, and the data plotted against PC2 would lie on a sharp, straight line, because they would be correlated perfectly with themselves. The colour of a galaxy is evidently composed of two components: a

`systematic' component correlated with all the other observables, and a `random' or `rogue' component correlated with nothing but itself. Appreciation of this non-intuitive point was greatly assisted by examining a PCA of simulated data (see Appendix A).

# Appendix A

HOW CORRELATIONS AND PRINCIPAL COMPONENTS RELATE;

ILLUSTRATED BY SIMPLE SIMULATIONS

To illustrate how correlated and uncorrelated data are represented in a principal component analysis (PCA), we give here the results of PCAs performed on correlation matrices for 200 samples of simulated data ($x$, $y$, $z$). We do so because we find the results to some extent non-intuitive and useful for interpreting Figure 4 in the main text. We consider three different cases: (i) $x$, $y$, and $z$ are all strongly-correlated with one another (correlation coefficient 0.8 for each pair); (ii) $x$ and $y$ are strongly-correlated with one another (correlation coefficient 0.8), but are uncorrelated with $z$; (iii) $x$ and $y$ are strongly-correlated with one another (correlation coefficient 0.8), but are weakly-correlated with $z$ (correlation coefficient 0.3 with each). Scatter plots showing the correlations between the simulated data, and between the data and the principal components are given in Figures A-1 and A-2 for case (i), Figures A-3 and A-4 for case (ii), and Figures A-5 and A-6 for case (iii). Note that case (iii) is very similar to the actual measured galaxy data discussed in the main text, where variable $z$ plays the role of colour, and Figure A-6 is instructive when compared with Figure 4 in the main text. It suggests that colour is a two component property, with one 'systematic' component weakly correlated with the other observables, and a stronger 'rogue' component correlated with nothing but itself.

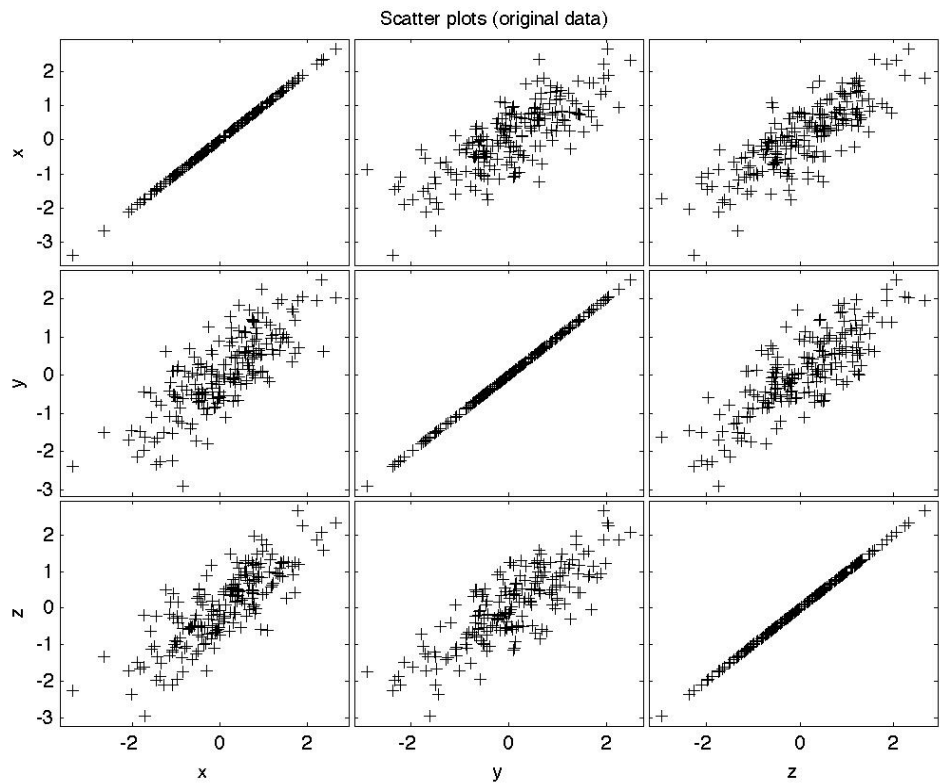

Figure A-1: Scatter plots showing the correlations between 200 samples of simulated data for case (i), where *x*, *y*, and *z* are all strongly-correlated with one another. The simulated data were drawn from a multivariate Gaussian distribution with correlation matrix having correlation coefficients of 0.8 for *x* and *y*, *y* and *z*, and *x* and *z*.

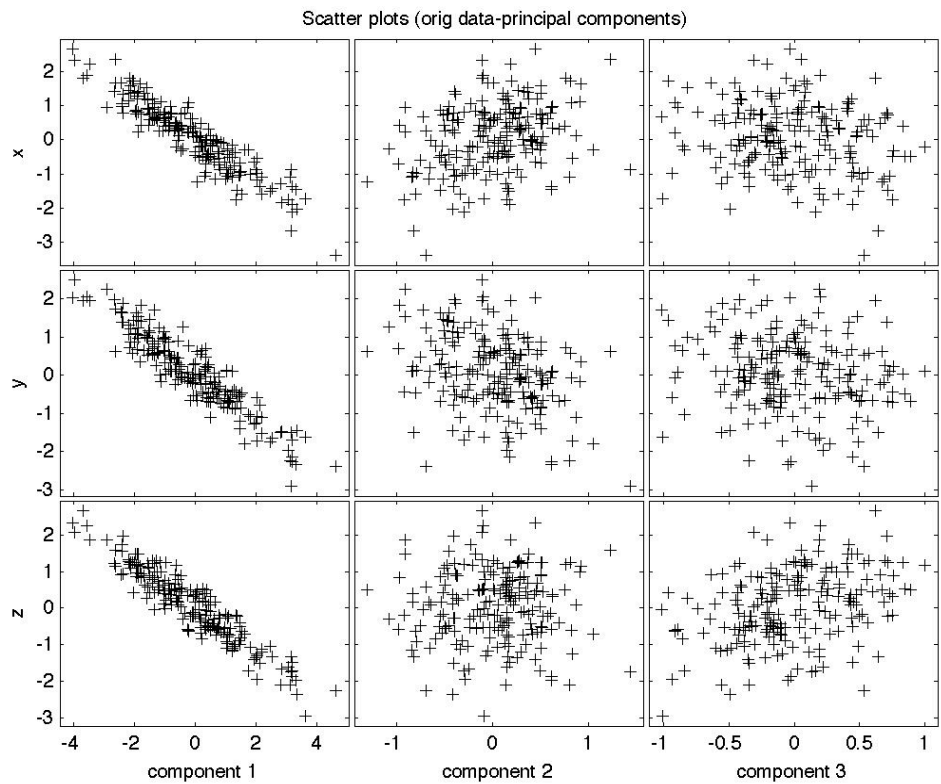

Figure A-2: Scatter plots showing correlations between the 200 samples of simulated data and the principal components for case (i), where $x$, $y$, and $z$ are all strongly-correlated with one another. For this case there is only one dominant principal component, PC1, with eigenvalue 2.6. (For perfect correlations this value would be 3,1 for each observable.) The eigenvalues of PC2 and PC3 are 0.22 and 0.18.) The direction cosines for PC1 relative to $x$, $y$, and $z$ are -0.57, -0.57, and -0.58, indicating that PC1 is an equally-weighted combination of the three variables. The correlation coefficients of the scatter plots in this figure and in other figures like it (e.g., Figures A-4 and A-6) are given by the product of the direction-cosines and the square-root of the PC eigenvalues.

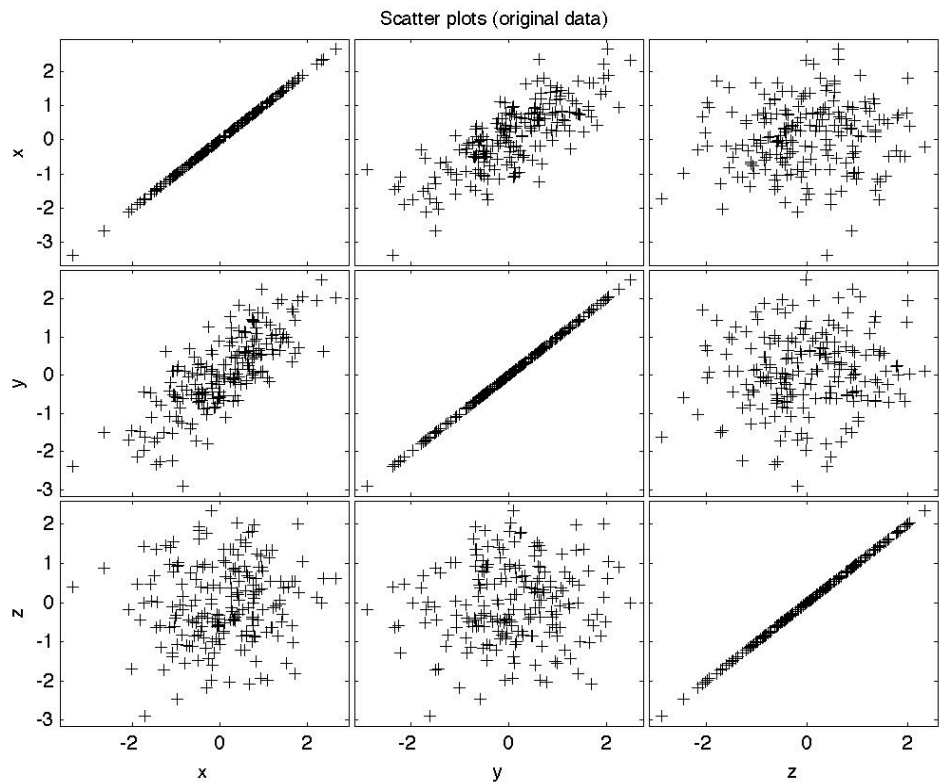

Figure A-3: Same as Figure A-1, but for case (ii), where *x* and *y* are strongly-correlated with one another, but are uncorrelated with *z*. The simulated data were drawn from a multivariate Gaussian distribution with a correlation matrix having correlation coefficient 0.8 for *x* and *y*, and zero correlation coefficient for both *x* and *z*, and *y* and *z*.

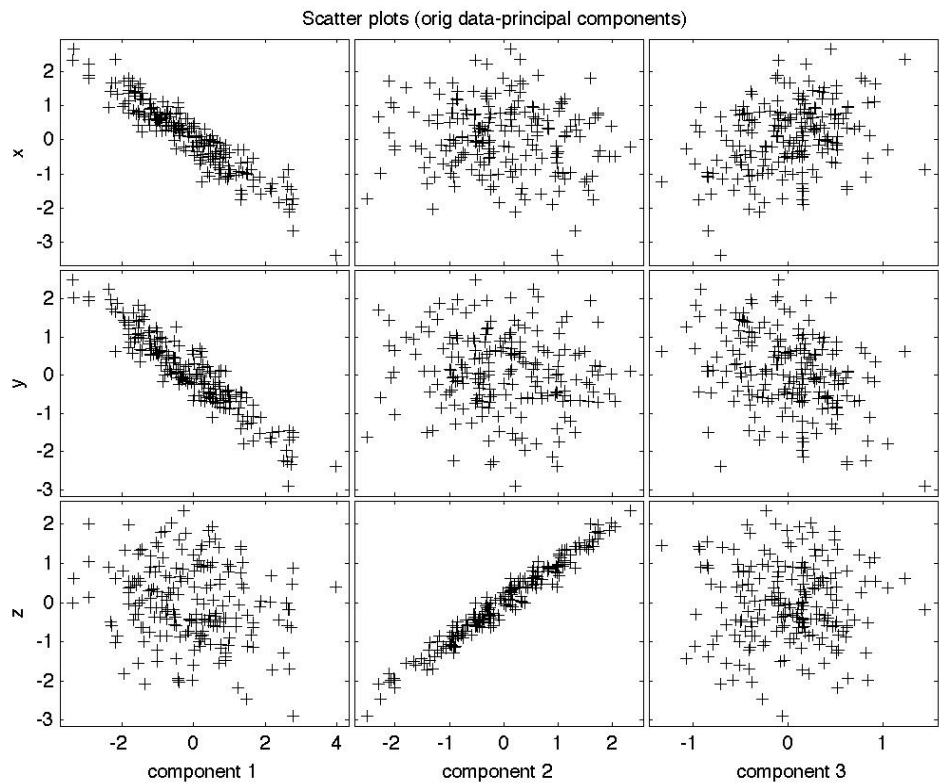

Figure A-4: Same as Figure A-2, but for case (ii), where *x* and *y* are strongly-correlated with one another, but are uncorrelated with *z*. Now there are two significant principal components, PC1 and PC2, with eigenvalues 1.79 and 0.98, respectively. (The eigenvalue of PC3 is 0.22.) The direction cosines for PC1 relative to *x*, *y*, and *z* are -0.70, -0.70, and -0.15. The direction cosines for PC2 are -0.10, -0.11, and 0.99, indicating that PC2 is just the variable *z*, which for this case is independent of the other variables, but of course is perfectly correlated with itself – hence PC2. For a random variable such as *z* to lead to a very strong principal component PC2 is not intuitive or at least it was not so to us.

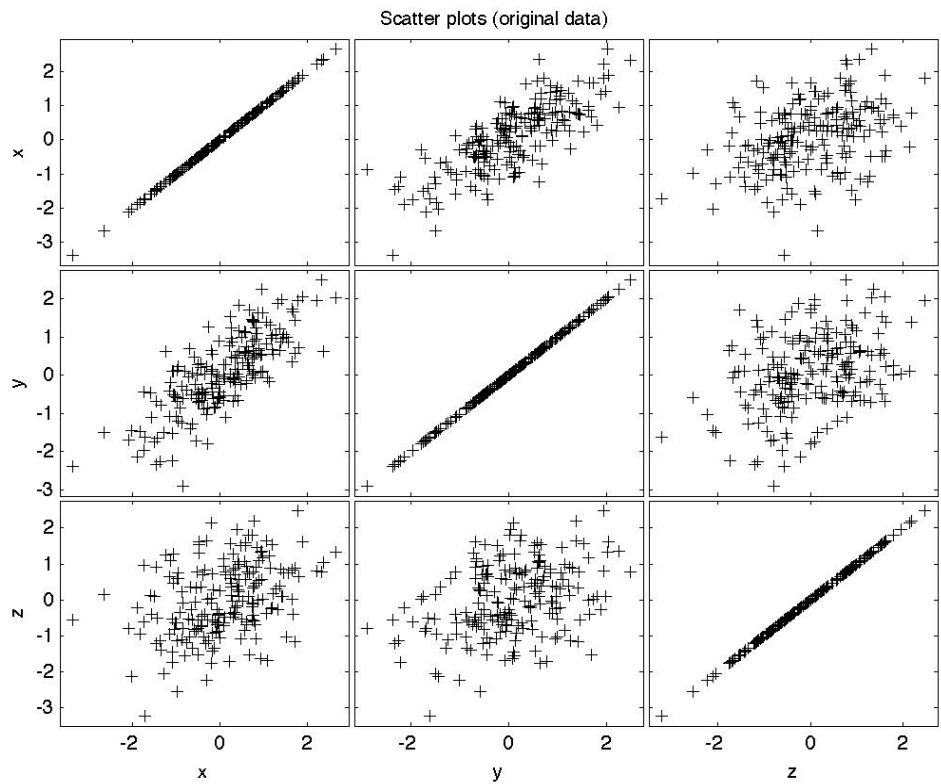

Figure A-5: Same as Figure A-3, but for case (iii), where $x$ and $y$ are strongly-correlated with one another but are weakly-correlated with $z$. The simulated data were drawn from a multivariate Gaussian distribution with a correlation matrix having correlation coefficient 0.8 for $x$ and $y$, and correlation coefficient 0.3 for both $x$ and $z$, and $y$ and $z$.

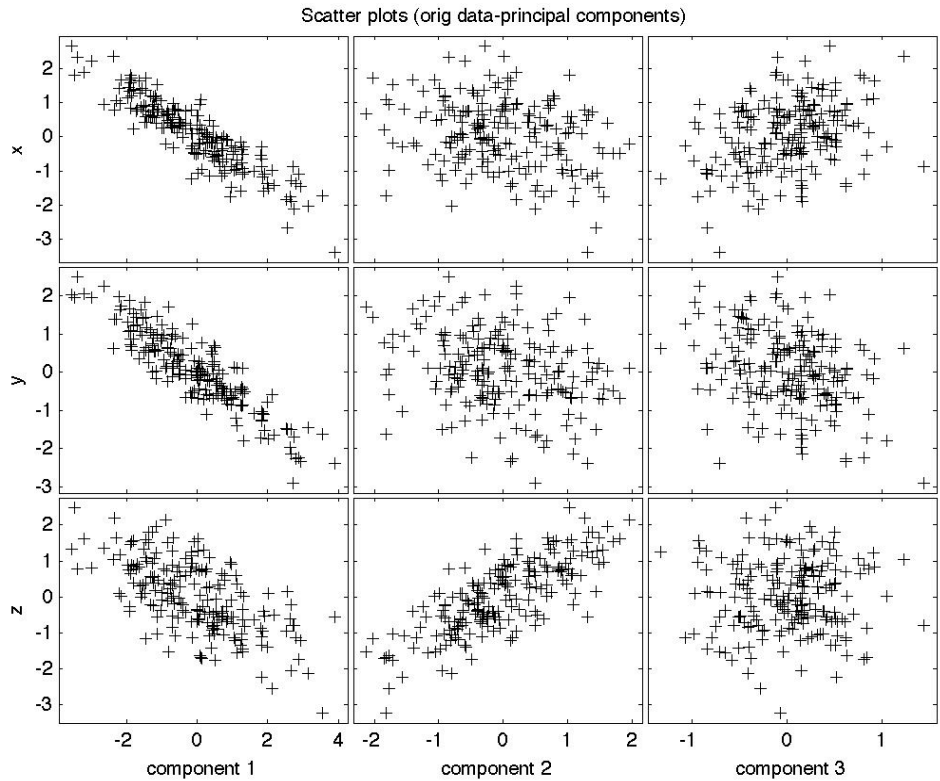

Scatter plots (orig data-principal components)

Figure A-6: Same as Figure A-4, but for case (iii), where *x* and *y* are strongly-correlated with one another but are weakly-correlated with *z*. For this case there is a dominant principal component, PC1, with eigenvalue 2.04, and a nominally significant second principal component, PC2, with eigenvalue 0.74. (The eigenvalue of PC3 is 0.22.) The direction cosines for PC1 relative to *x*, *y*, and *z* are -0.63, -0.63, and -0.45, respectively, indicating that it has components along all three variables. The direction cosines for PC2 are -0.31, -0.32, and 0.89, indicating that it is primarily composed of the variable *z*, which for this case is weakly-correlated with *x* and *y*. This simulation is very similar to the actual measured galaxy data discussed in the main text, where variable *z* plays the role of colour. This suggests that colour has two components; one weakly correlated with the other observables, and another stronger 'rogue' component which is correlated with nothing but itself.

# Appendix B

RESCALED FIGURES

The figures in the main text are for the data *normalized* by their standard deviations. The following are the same figures, but expressed in terms of the un-normalized data, which are easier to interpret directly, and are given in solar masses, solar luminosities, and kiloparsecs, as appropriate. None of the figures themselves are changed, nor the science – e.g., the eigenvalues. (The standard deviations of the data, which allow one to go back-and-forth between the normalized and un-normalized data are 0.3027, 0.2931, 0.7441, 0.5564, 0.7166, and 0.1727 for $\log R_{50}$, $\log R_{90}$, $\log L_g$, $\log M_{HI}$, $\log M_d$, and colour, respectively.)

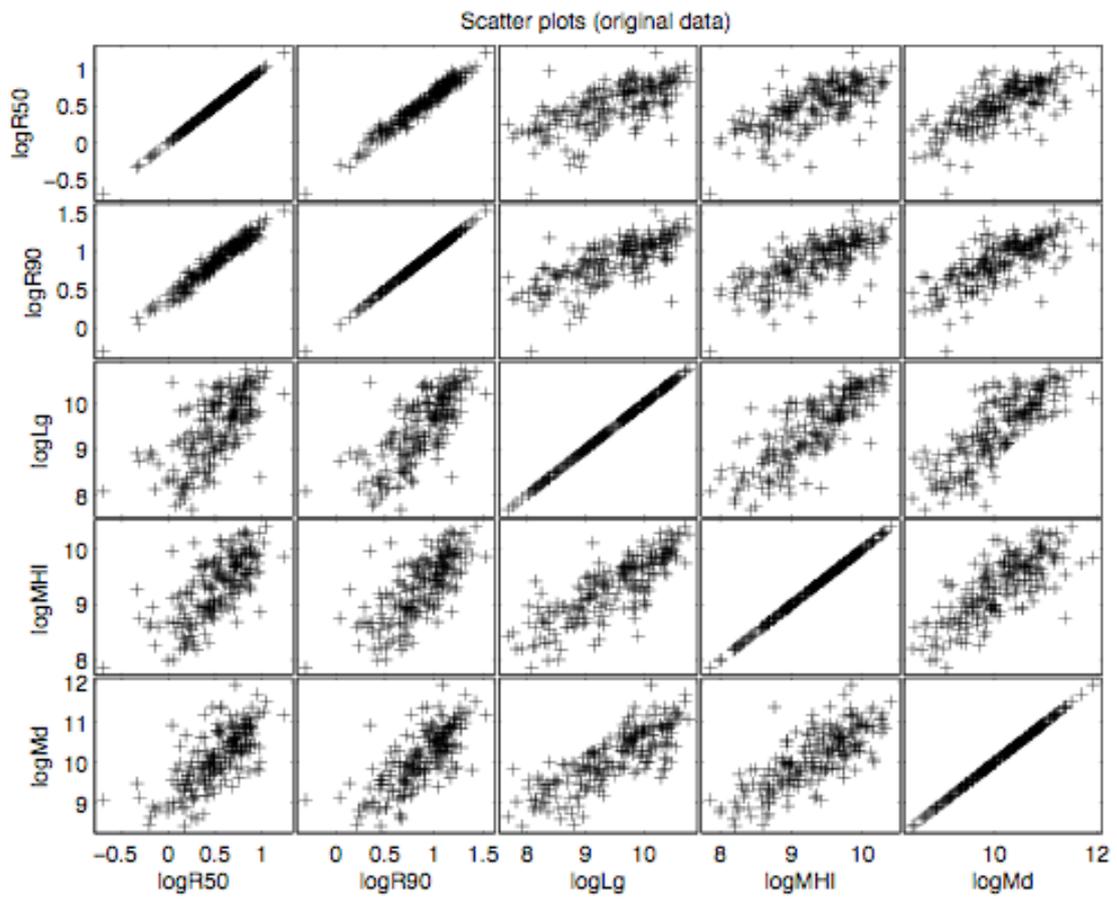

Figure 1

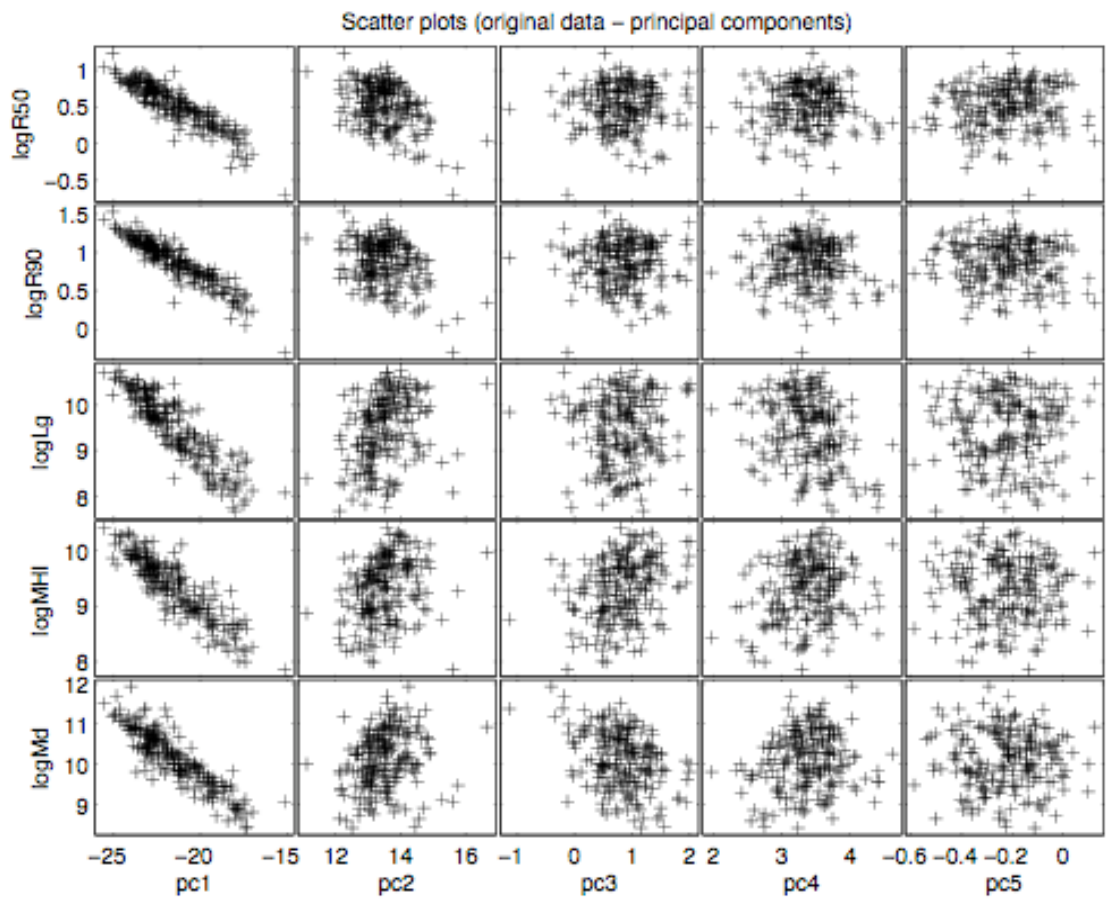

Figure 2

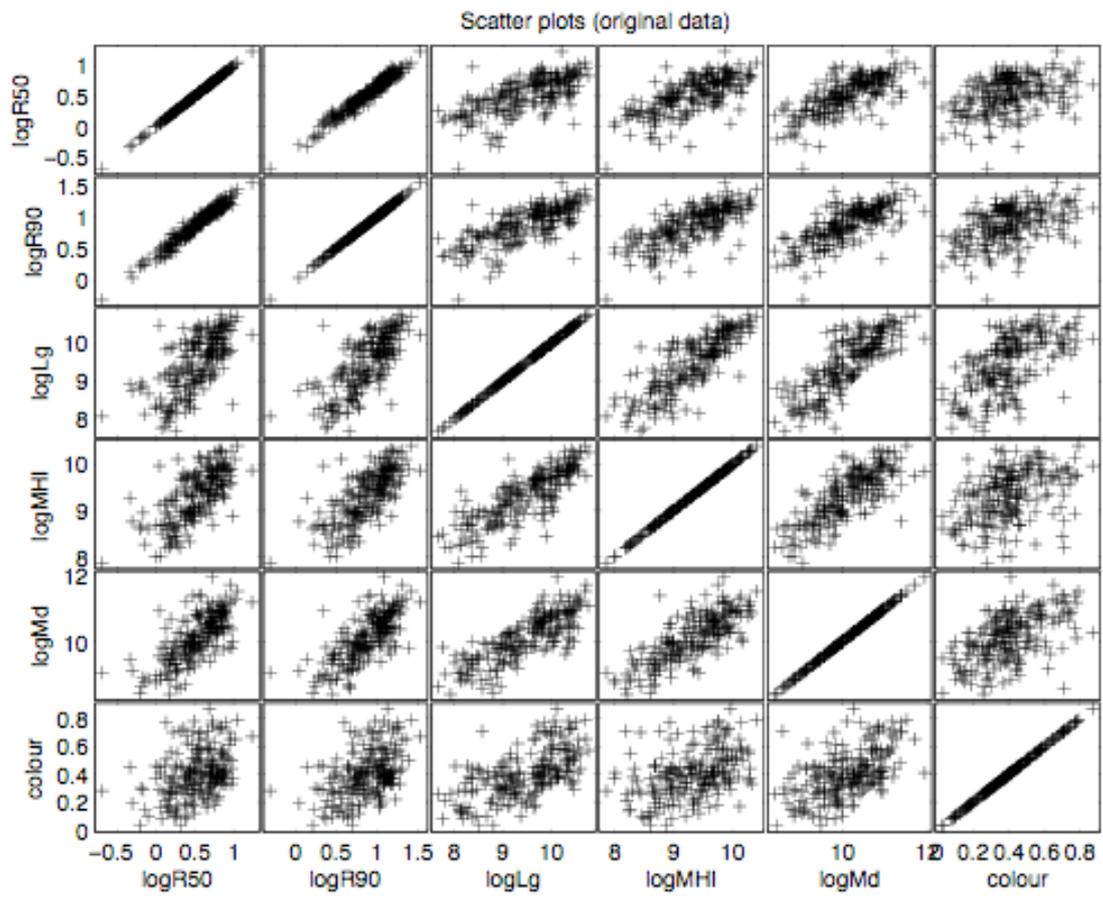

Figure 3

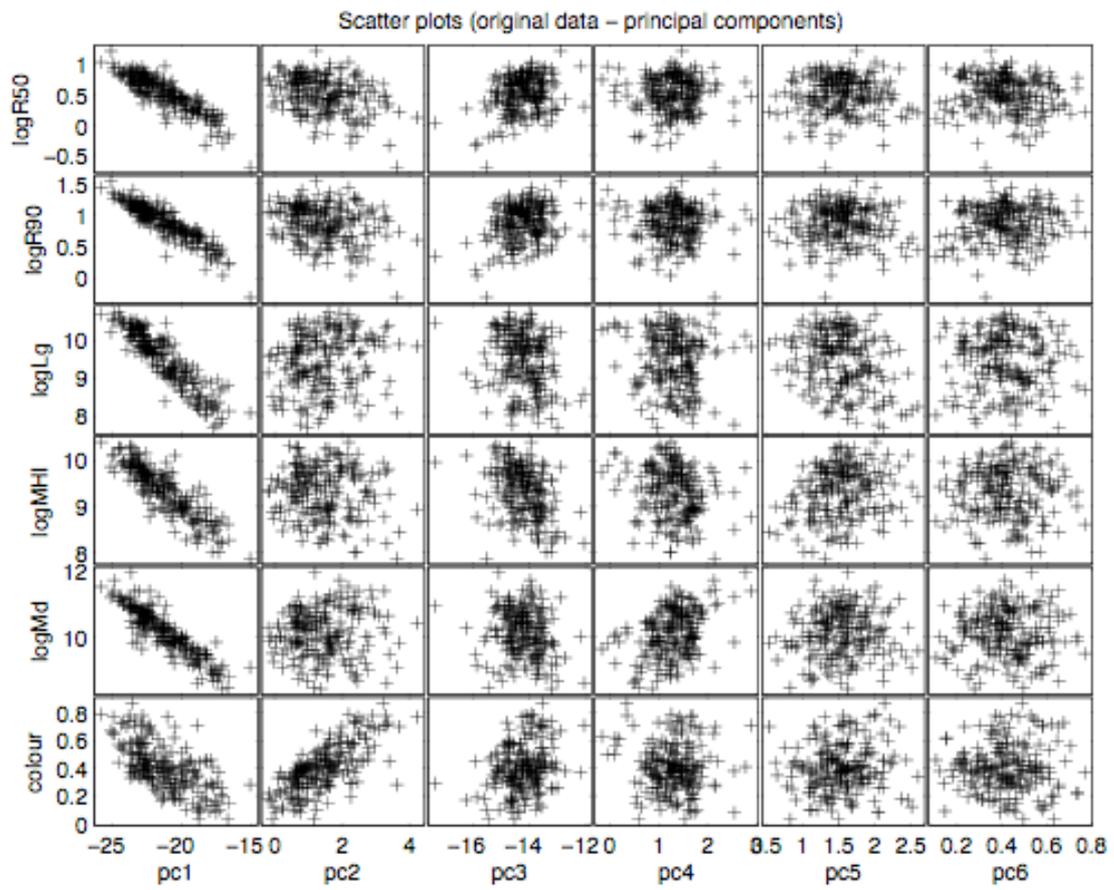

Figure 4